\documentclass[twocolumn,runningheads]{svjour2}
\smartqed  
\usepackage{graphicx}
\usepackage{mathptmx}   
\usepackage[authoryear]{natbib}
\newcommand{\noprintlabel}{}
\newcommand{\etal}{et al.}
\newcommand{\fig}{.}
\newcommand{\eplus}{{\rm e^+}}

\newcommand{\hyd}{{{}^{1}\rm H}}
\newcommand{\deut}{{{}^{2}\rm D}}
\newcommand{\helthree}{{{}^{3}\rm He}}
\newcommand{\helfour}{{{}^{4}\rm He}}
\newcommand{\berseven}{{{}^{7}\rm Be}}

\newcommand{\litseven}{{{}^{7}\rm Li}}

\newcommand{\cartwelve}{{{}^{12}\rm C}}

\newcommand{\nitfourteen}{{{}^{14}\rm N}}
\newcommand{\nitfifteen}{{{}^{15}\rm N}}
\newcommand{\oxyfifteen}{{{}^{15}\rm O}}
\newcommand{\oxysixteen}{{{}^{16}\rm O}}
\newcommand{\oxyseventeen}{{{}^{17}\rm O}}
\newcommand{\floseventeen}{{{}^{17}\rm F}}
\newcommand{\nablarad}{\nabla_{\rm rad}}
\newcommand{\nablaad}{\nabla_{\rm ad}}
\def\biblab #1{\ifx\noprintlabel\undefined{\bf [#1]}\fi}
\newcommand{\logten}{\log_{10}}
\newcommand{\CC}{{\cal C}}
\newcommand{\CD}{{\cal D}}
\newcommand{\CM}{{\cal M}}
\newcommand{\CR}{{\cal R}}
\newcommand{\CV}{{\cal V}}
\newcommand{\dd}{{\rm d}}
\newcommand{\clight}{\tilde c}
\newcommand{\KA}{K_{\rm A}}
\newcommand{\KB}{K_{\rm B}}
\newcommand{\Msun}{{\rm M}_{\odot}}
\journalname{Astrophysics and Space Science (CoRoT/ESTA Volume)}
\begin{document}

\title{ASTEC -- the Aarhus STellar Evolution Code
}


\author{J{\o}rgen Christensen-Dalsgaard
}


\institute{J{\o}rgen Christensen-Dalsgaard \at
              Institut for Fysik og Astronomi, og Dansk AsteroSeismisk Center \\
              Bygning 1520 \\
              Aarhus Universitet \\
              DK-8000 Aarhus C \\
              Denmark \\
              Tel.: +45 89 42 36 14\\
              Fax: +45 86 12 07 40\\
              \email{jcd@phys.au.dk}           
}

\date{Received: date / Accepted: date}

\maketitle
\begin{abstract}
The Aarhus code is the result of a long development, starting in 1974,
and still ongoing.
A novel feature is the integration of the computation of adiabatic oscillations
for specified models as part of the code.
It offers substantial flexibility in terms of microphysics and has been
carefully tested for the computation of solar models.
However, considerable development is still required in the treatment of
nuclear reactions, diffusion and convective mixing.
\end{abstract}

\section{Introduction}\label{intro}

What has become ASTEC started its development in Cambridge around 1974.
The initial goal was to provide an improved equilibrium model for
investigations of solar stability, following earlier work by
\citet{Christ1974}.
However, with the initial evidence for solar oscillations and the prospects
for helioseismology \citep{Christ1976}
the goals were soon extended to provide models for comparison with the
observed frequencies.
Given the expected accuracy of these frequencies, and the need to use them
to uncover subtle features of the solar interior, more emphasis was placed
on numerical accuracy than was perhaps common at the time.
The code drew some inspiration from the Eggleton stellar evolution code
\citep{Egglet1971},
which had been used previously, but the development was fully independent
of that code.
An early description of the code was given by 
\citet{Christ1982},
with further extensive details provided by \citet{Christ1978};
many aspects of this still stand and will only be summarized here.
With the increasing quality and extent of the helioseismic data the code
was further developed, to allow for more realistic physics; a major improvement
was the inclusion of diffusion and settling \citep{Christ1993}.
This led to the so-called Model~S of the Sun \citep{Christ1996}
which has found extensive use as reference for helioseismic investigations
and which at the time provided reasonably up-to-date representations of
the physics of the solar interior.
In parallel, extensions have been made to the code to consider the evolution
of stars other than the Sun; these include the treatment of convective cores
and core overshoot, attempts to model red-giant evolution and the inclusion
of helium burning.
This development is still very much under way.

For use in asteroseismic fitting a version of the code has also been
developed in the form of a subroutine with a reasonably simple calling
structure which also includes the computation of adiabatic
oscillation frequencies as part of the computation.
The combined package is available as a single tar file, making the installation
relatively straightforward, and the code has been successfully
implemented on a variety of platforms.
Nevertheless, it is sufficiently complex that a general release is probably
not advisable.

\section{Equations and numerical scheme}\label{sec:eqnum}


\subsection{Formulation of the equations}\label{sec:eq}

The equations of stellar structure and evolution are written with 
$x = \logten q$ as independent variable, where $\logten$ is logarithm to
base 10; here $q = m/M$ is the mass fraction where $m$ is the mass interior
to the point considered and $M$ is the photospheric
mass of the star, defining the photosphere as the point where the temperature
is equal to the effective temperature.
The equations are written on the form
\begin{eqnarray}
{\partial \logten r \over \partial x} & = & {M \over 4 \pi \rho} {q \over r^3} 
\; ,
\label{eq:radius} \\
{\partial \logten p \over \partial x} & = &
- {G M^2 \over 4 \pi} {q^2 \over r^4 p} \; ,
\label{eq:hydro} \\
{\partial \logten T \over \partial x} & = & 
\nabla {\partial \logten p \over \partial x} \; ,
\label{eq:temp} \\
{\partial \logten L \over \partial x} & = & 
M \left(\epsilon - {\partial H \over \partial t} 
+ {1 \over \rho} {\partial p \over \partial t} \right) {q \over L} \; ,
\label{eq:energy} \\
{\partial X_k \over \partial t } & = & \CR_k +
{\partial \over \partial m} \left( \CD_k {\partial X_k \over \partial m} \right)
+ {\partial \over \partial m} ( \CV_k X_k ) \; 
\label{eq:dif} \\
&& k = 1, \ldots, K \nonumber \; .
\end{eqnarray}
Here $r$ is distance to the centre, $\rho$ is density, $p$ is pressure,
$G$ is the gravitational constant, $T$ is temperature, 
$L$ is the luminosity at $r$,
$\epsilon$ is the energy generation rate per unit mass, $H$ is enthalpy per
unit mass, $X_k$ is the mass fraction of element $k$, $\CR_k$ is the rate
of change of $X_k$ due to nuclear reactions, and $\CD_k$ and $\CV_k$ are
the diffusion and settling coefficients for element $k$.
Also, the value of $\nabla = \dd \log T / \dd \log p$ depends on whether
the region is convectively stable or not.
In the case of a convectively stable region,
\begin{equation}
\nabla = \nablarad
\equiv \psi {3 \over 16 \pi a \clight} {\kappa \over T^4} {L p \over G m} 
\; ,
\label{eq:radgrad}
\end{equation}
where $a$ is the radiation density constant, $\clight$ is the speed of
light, and $\kappa$ is the opacity.
The factor $\psi$ is discussed following Eq.~\ref{eq:ttau} below.
The calculation of the temperature gradient in convective regions is
discussed in Section~\ref{sec:conv}.
In Eq.~\ref{eq:dif} the derivatives with respect to $m$ should obviously
be expressed in terms of derivatives with respect to $x$.

As discussed below, it is convenient to introduce
the diffusion velocity on the form
\begin{equation}
Y_k = \CD_k {\partial X_k \over \partial m} + \CV_k X_k \; ;
\label{eq:difvel}
\end{equation}
then Eqs~\ref{eq:dif} can be written as
\begin{eqnarray}
{\partial X_k \over \partial x} & = &
{\CM M q \over \CD_k} (Y_k - \CV_k X_k) \; , 
\label{eq:difx} \\
{\partial Y_k \over \partial x} & = &
\CM M q  \left({\partial X_k \over \partial t} - \CR_k \right) \; .
\label{eq:dify}
\end{eqnarray}
Here $\CM = \log 10$, $\log$ being the natural logarithm.
(In the code $\tilde Y_k = M^{-1} Y_k$ is used as variable.)
For elements where diffusion and settling are ignored the original form,
Eq.~\ref{eq:dif}, with $\CD_k$ and $\CV_k$ set to zero is obviously used.

\subsection{Boundary conditions}

The centre, $r = 0$, is obviously a singular point of Eqs \ref{eq:radius} --
\ref{eq:energy}.
As discussed in considerable detail by \citet{Christ1982},
this is treated by expanding the variables to second significant order 
in $r$ to obtain the required conditions at the innermost point in the
numerical solution.
These are written in the form of expressions of the $m$ and $L$
at the innermost mesh point in terms of the other variables.
The treatment of diffusion and settling has yet to be included consistently
in this expansion;
currently, conditions are imposed that essentially set the $\tilde Y_k$
to zero at the innermost meshpoint.
The outer boundary of the solution of the full evolution equations is taken
to be at the photosphere, defined as the point where $T = T_{\rm eff}$,
the effective temperature.
Consequently an obvious boundary condition is
\begin{equation}
L = 4 \pi r^2 \sigma T^4 \; ,
\end{equation}
where $\sigma$ is the Stefan-Boltzmann constant.
To obtain a second condition, a expression is assumed for the dependence
of $T$ on optical depth $\tau$, conveniently written as
\begin{equation}
T = T_{\rm eff} [\tau + q(\tau)]^{1/4} \; .
\label{eq:ttau}
\end{equation}
Based on this the equation of hydrostatic equilibrium, on the form
\begin{equation}
{\dd p \over \dd \tau} = {g \over \kappa} \; ,
\label{eq:atmohydro}
\end{equation}
can be integrated assuming, currently, a constant gravitational acceleration
$g$.
The surface boundary condition 
\begin{equation}
p = {2 \tau g \over \kappa} \; ,
\end{equation}
based on an approximate treatment of the outer parts of the atmosphere,
is applied at the top of the atmosphere, $\tau = \tau_{\rm min}$,
typically taken to be $10^{-4}$.
The boundary condition is obtained by equating the pressure resulting 
from integrating Eq.~\ref{eq:atmohydro} to the pressure 
in the interior solution.

The $T(\tau)$ relation and Eq.~\ref{eq:atmohydro}
obviously define the temperature gradient $\nabla$ in the atmosphere.
To ensure a continuous transition to the interior I follow
\citet{Henyey1965} and include the factor $\psi = 1 + q'(\tau)$ 
in Eq.~\ref{eq:radgrad}.

\subsection{Numerical scheme}

As indicated in Eqs \ref{eq:radius} -- \ref{eq:dif}
and \ref{eq:difx}, \ref{eq:dify}, the dependent variables on the left-hand
sides of the equations are expressed in terms of the set
\begin{eqnarray}
\{y_i\} &=& \{\logten r, \logten p, , \logten T, \logten L, X_k, 
\tilde Y_{k'} \} \; ,
\nonumber \\
&&\qquad i = 1, \ldots, I \; .
\end{eqnarray}
Here the index $k$ runs over all the abundances considered,
whereas the index $k'$ runs over those elements for which diffusion is included.
In the solution of the equations, the computation of the right-hand sides
is the heaviest part of the calculation and hence needs to be optimized in
terms of efficiency.
As discussed below, this may involve expressing the thermodynamic state in
terms of a different set of variables $\{z_j\}$,
related to the $\{y_i\}$ by a non-singular transformation.
Using also the reformulation, Eqs \ref{eq:difx} and \ref{eq:dify},
of the diffusion equation, the combined set of
stellar-evolution equations 
consists of equations of one of the following three types:
\begin{eqnarray}
\label{eq:typei}
{\partial y_l \over \partial x} & =& f_l(x; z_i; t) \; ,
\qquad l = 1, \ldots I_1 \qquad \hbox{\bf Type I} \\
\label{eq:typeii}
{\partial y_p \over \partial x} & = & f_p(x; z_i; t) +
\sum_{i = 1}^I \Lambda_{pi} (x; z_j; t) {\partial y_i \over \partial t} \; ,
\\
&& p = I_1 +1, \ldots I_1 + I_2  \qquad \hbox{\bf Type II}  \\
\label{eq:typeiii}
{\partial y_u \over \partial t} & =& f_u(x; z_i; t)  \; , \\
&& u = I_1 + I_2 +1, \ldots I_1 + I_2 + I_3  \qquad \hbox{\bf Type III} 
\end{eqnarray}
The equations of Type III are obviously relevant to the evolution of abundances
of elements for which diffusion is not taken into account.
To this must be added the transformation
\begin{equation}
y_i = y_i(x; z_j; t) \; , \qquad i = 1, \ldots I \; .
\end{equation}
In practice, $z_i = y_i$ for $i \neq 2$; 
the choice of $z_2$ depends on the specific equation of state considered.
The equations are solved on the interval $[x_1, x_2]$, with boundary
conditions that can be expressed as
\begin{eqnarray}
g_{\alpha} ( x_1 ; z_i ( x_1 ) ) &=& 0  \; ,   \qquad
\alpha = 1,  \ldots ,  \KA \; , \label{eq:bca} \\
g_{\beta} ( x_2 ;  z_i ( x_2 ) ) &=& 0  \; , \qquad
\beta = \KA + 1,  \ldots ,  \KA + \KB \; .
\label{eq:bcb}
\end{eqnarray}

The equations are solved by means of the Newton-Raphson-Kantorovich scheme
\citep[see][]{Richtm1957, Henric1962},
in the stellar-evolution context known as the Henyey scheme 
\citep[e.g.,][]{Henyey1959, Henyey1964}.
We introduce a mesh $x_1 = x^1 < x^2 < \ldots x^N = x_2$ in $x$ and consider
two timesteps $t^s$ and $t^{s+1}$, where the solution is assumed to be known
at timestep $t^s$.
Also, we introduce
\begin{eqnarray}
&& y_i^{n,s} = y_i(x^n, t^s) \; , \quad  
z_i^{n,s} = z_i(x^n, t^s) \; , \nonumber \\
&& f_i^{n,s} = f_i(x^n; z_j^{n,s}; t^s) \; , 
\label{eq:defn}
\end{eqnarray}
with a similar notation for $\Lambda_i^{n,s}$, as well as for quantities at
timestep $t^{s+1}$.
Then Eqs \ref{eq:typei} -- \ref{eq:typeiii} are replaced by the following 
difference equations:
\begin{eqnarray}
&& y_l^{n+1,s+1} - y_l^{n,s+1} 
= {1 \over 2} \Delta x^n 
(f_l^{n+1,s+1} + f_l^{n,s+1}) \; , \nonumber \label{eq:dtypei} \\
&& \qquad n = 1, \ldots, N-1 \; ; \quad l = 1, \ldots I_1 \; ,\\
&& \theta_p (y_p^{n+1,s+1} - y_p^{n,s+1}) +
(1 - \theta_p) (y_p^{n+1,s} - y_p^{n,s}) \nonumber \\
&& = {1 \over 2} \Delta x^n 
\LARGE\{ \theta_p (f_p^{n+1,s+1} + f_p^{n,s+1}) \nonumber \\
&& + (1 - \theta_p) (f_p^{n+1,s} + f_p^{n,s}) \nonumber \\
&& + \sum_{i=1}^I \left[ \theta_p \Lambda_{pi}^{n+1,s+1} +
(1 - \theta_p) \Lambda_{pi}^{n+1,s}\right] 
(z_i^{n+1,s+1}-z_i^{n+1,s})/\Delta t^s
\nonumber \\
&& + \sum_{i=1}^I \left[ \theta_p \Lambda_{pi}^{n,s+1} +
(1 - \theta_p) \Lambda_{pi}^{n,s}\right] (z_i^{n,s+1}-z_i^{n,s})/\Delta t^s
\LARGE\}
\; , \nonumber \\
&& \qquad n = 1, \ldots, N-1 \; ; \quad p = I_1+1, \ldots I_1 +I_2 \; ,
\label{eq:dtypeii} \\
&& y_u^{n,s+1} - y_u^{n,s} = \Delta t^s \left[ \theta_u f_u^{n,s+1}
+ (1 - \theta_u) f_u^{n,s} \right] \; , \nonumber \\
&& \qquad n = 1, \ldots, N \; ; \quad u = I_1 +I_2+1, \ldots I_1+I_2+I_3 \; .
\label{eq:dtypeiii}
\end{eqnarray}
Here $\Delta x^n = x^{n+1} - x^n$ and $\Delta t^s = t^{s+1} - t^s$.
Also, the parameters $\theta_i$ allow setting the centralization of
the difference scheme in time, $\theta_i = 1/2$ corresponding to 
time-centred differences and $\theta_i = 1$ to a fully implicit scheme.
The former clearly provides higher precision but potentially less 
stability than the latter;
thus time-centred differences are typically used for processes occurring
on a slow timescale, such as the change in the hydrogen abundance, whereas
the implicit scheme is used, e.g., for the time derivatives in the energy
equation where the characteristic timescale is much shorter than the
evolution time and hence short compared with 
the typical step $\Delta t$ in time.

The algebraic equations \ref{eq:dtypei} -- \ref{eq:dtypeiii},
together with the 
boundary conditions, Eqs~\ref{eq:bca} and \ref{eq:bcb}, are solved
using Newton-Raphson iteration, to determine the solution
$\{z_i^{n, s+1}\}$ at the new time step \citep[see also][]{Christ1982}.
Given a trial solution 
$\{\bar z_i^{n, s+1}\}$ the equations are linearized 
in terms of corrections 
$\delta z_i^{n, s+1} = z_i^{n, s+1} -\bar z_i^{n, s+1}$ and the
resulting linear equations are solved using forwards elimination
and backsubstitution \citep[e.g.,][]{Baker1971}.
This process is repeated with the thus corrected solution as trial, until
convergence.


The initial ZAMS model is assumed to be static and with a prescribed
chemical composition.
Thus in this case only Eqs~\ref{eq:dtypei} are solved.
Time evolution is started with a very small $\Delta t$ for the initial
non-zero timestep.

The number $N$ of meshpoints is kept fixed during the evolution,
but the distribution of points is varied according to the change in
the structure of the model.
The mesh algorithm is based on the first-derivative stretching 
introduced by \citet{Gough1975} \citep[see also][]{Egglet1971},
taking into account the variation of several relevant variables.
An early version of the implementation was described by \citet{Christ1982},
but this has subsequently been substantially extended and is still
under development.
In particular, a dense distribution of points is set up near the boundaries
of convection zones, although so far points are not adjusted so as to be
exactly at the edges of the zones.
After completing the solution at each timestep the new mesh is determined
and the model transferred to this mesh using, in general,
third-order interpolation;
linear interpolation is used
near boundaries of convective regions and in other regions where the
variation of the variables is not smooth.

Having computed model at timestep $t^{s+1}$ the next time\-step is determined
from the change in the model between timesteps $t^s$ and $t^{s+1}$;
this involves a fairly complex algorithm limiting the change in, e.g.,
$\logten p$, $\logten T$ and $X$ at fixed $m$.
To compensate for the fairly crude treatment of the composition in a
possible convective core, the change of $X$ in such a core is given higher
weight than the general change in $X$.
The algorithm correctly ensures that very short steps in time are taken in
rapid evolutionary phases.
In typical simple calculations,
assuming $\helthree$ to be in nuclear equilibrium,
roughly 35 (100) timesteps are needed to reach the end
of central hydrogen burning in models without (with) a convective core,
and 100 (200) steps to reach the base of the red-giant branch.
Calculations with more complex physics or requiring higher numerical
precision obviously require a substantially higher number of timesteps.
Evolution up the red giant branch typically requires a large number 
of timesteps owing to the rapid changes at fixed $m$ in the hydrogen-burning
shell%
\footnote{This problem is avoided in the implementation by
\citet{Egglet1971} where the equations are solved using an independent variable 
that is directly tied to such strong variations in the model.}
although the timestep algorithm has options to reduce the weight given
to this region.





\section{Microphysics}\label{sec:microphys}

As the code has developed over the years a number of options have been 
included for the microphysics, although not all of these have been kept 
up to date or properly verified.
As a general principle, the code has been written in a modular way, so that
replacing, for example, routines for equation of state or opacity has been
relatively straightforward.
It should be noted that the use of a different set $\{z_i\}$ of dependent
variables on the right-hand side of the equations yields additional
flexibility in the replacement of aspects of the physics.
\citet{Dappen2000} provided a review of the treatment of the equation of
state and opacity in stellar modelling.
A review of nuclear reactions in the solar interior, of 
relevance in the more general stellar case, was provided by
\citet{Parker1991}.

\subsection{Equation of state}
\label{sec:eos}

The original version of the code used the \citet{Egglet1973} equation of state 
and that remains an often used option.
In this case $z_2 = \logten f$,
where $f$, introduced by Eggleton et al. 
and related to the electron-degeneracy parameter,
is used as one of the thermodynamic variables;
this allows explicit calculation of partial ionization and hence a very 
efficient evaluation of the required thermodynamic quantities.
The formulation also includes a crude but thermodynamically consistent 
implementation of `pressure ionization' (which actually results mainly as a
result of the high density in deep stellar interiors) which provides apparently
reasonable results.
The fact that the whole calculation is done explicitly makes it entirely
feasible, if somewhat cumbersome, to evaluate analytical derivatives.
Up to second derivatives of pressure, density and enthalpy
are provided in a fully consistent manner, whereas third derivatives, required
for the central boundary condition, assume full ionization.
Partial electron degeneracy is included in the form of expansions that cover
all levels of degeneracy and relativistic effects.

As an upgrade to the basic EFF formulation \citet{Christ1992}
included Coulomb effects, in the Debye-H\"uckel formulation,
following \citet{Mihala1988};
unlike some earlier treatments of the Coulomb effect this ensured that 
thermodynamic consistency was maintained; as a result, a substantial
effect was found also on the degrees of ionization of hydrogen and helium,
resulting from the change in the chemical potential.
The computation of the Coulomb effects consequently requires some iteration,
even if the EFF variables are used, and hence some increase in computing time.
However, the resulting equation of state captures major aspects of 
the more complex forms discussed below.

More realistic descriptions of the equation of state require computations that 
are currently too complex to be included directly in stellar evolution
calculations.
Thus interpolation in pre-computed tables is required.
The first such set to be included was the so-called MHD equation of state
\citep{Mihala1990},
based on the {\it chemical picture\/} where the thermodynamic state is
obtained through minimization of an expression for the free energy including
a number of contributions.
Application of this formulation to a comparison with observed solar oscillation
frequencies showed a very substantial improvement in the agreement between
the Sun and the model \citep{Christ1988}.
Further updates to the MHD equation of state have been made but they have
so far not been implemented in ASTEC.
An alternative description is provided by the so-called OPAL equation of state
\citep{Rogers1996},
based on the {\it physical picture\/} where the thermodynamic state is 
obtained from an activity expansion.
This has been the preferred equation of state for solar modelling
with ASTEC, used, e.g., in Model~S.

The early versions of both the MHD and OPAL equations of state suffered from
a neglect of relativistic effects in the treatment of the electrons
\citep{Elliot1998}.
This has since been corrected \citep{Gong2001, Rogers2002}.
Also, the OPAL formulation suffered from inconsistencies between some of
the variables provided \citep[e.g.][]{Boothr2003, Scufla2007};
effects of the inconsistency on the computation
of adiabatic oscillations are discussed by \citet{Moya2007}.
This has been improved in the latest version of the OPAL equation of state,%
\footnote{see {\tt \hfill\break
http://www-phys.llnl.gov/Research/OPAL/opal.html}}
which has also been implemented in ASTEC.

Interpolation in the OPAL tables uses quadratic interpolation in $\rho$, $T$
and $X$.
Typically, a single representative value of $Z$ is used, even in cases with
diffusion and settling of heavy elements, although the code has the option
of using linear interpolation between two sets of tables with different $Z$.


\subsection{Opacity}
\label{sec:opac}

The early treatment of the opacities used tables from 
\citet{Cox1970} and \citet{Cox1976}
with an interpolation scheme developed by \citet{Cline1974}.
A major improvement was the implementation of the OPAL opacity tables
\citep{Rogers1992, Iglesi1992}.
With various updates of the tables this has since been the basis for the
opacity calculation in the code.
The most recent tables, including a variety of mixtures of the heavy elements,
are based on the computations by \citet{Rogers1995}.
Atmospheric opacities must be supplied separately;
here tables by \citet{Kurucz1991}, \citet{Alexa1994} or \citet{Fergus2005}
have been 
used, with a smooth matching to the interior tables at $\logten T = 4$.
Electron conduction may be included based on the tables of \citet{Itoh1983}.



Interpolation in the OPAL tables is carried out with 
schemes developed by G. Houdek \citep[see][]{Houdek1993, Houdek1996}.
The tables are provided as functions of $(R, T)$, where $R = \rho/T^3$.
For interpolation in $(\log R, \log T)$ two schemes are available.
One uses a minimum-norm algorithm with interpolating function defined in
a piecewise fashion over triangles in the $\log R - \log T$ plane
\citep{Nielso1980, Nielso1983}.
The second scheme uses birational splines \citep{Spath1991}.
In practice the latter scheme has been used for most of the model calculations
with ASTEC.
Interpolation in $X$ and $\log Z$ is carried out using the univariate
scheme of \citet{Akima1970}.
Note that the relative composition of $Z$ is assumed to be fixed; thus
differential settling or changes in heavy-element composition resulting
from nuclear burning are not taken into account.


The code includes flexible ways of modifying the opacity, to allow tests of
the sensitivity of the model to such modifications.
An extensive survey of the response of solar models to localized opacity
changes, specified as functions of $\logten T$, was made by
\citet{Tripat1998}.

\subsection{Nuclear reactions}

The calculation of the nuclear reaction rates is based on the usual
approximations to the reaction integrals \citep[e.g.,][]{Clayto1968}
using a variety of coefficients 
\citep{Bahcal1995, Adelbe1998, Angulo1999}.
Electron screening is computed in the \citet{Salpet1954} approximation.
Electron capture by $\berseven$ is treated according to \citet{Bahcal1969}.

The nuclear network is relatively limited and is one of the points where the
code needs improvement.
This is to some extent a heritage of its origin as a solar-modelling code,
as well as a consequence of the fact that pre-main-sequence evolution is
not computed.
In the pp chains $\deut$, $\litseven$ and $\berseven$ 
are assumed to be in nuclear equilibrium.
On the other hand, the code has the option of
following the evolution of $\helthree$, although in many calculations
it is sufficient to assume $\helthree$ to be in nuclear equilibrium.
To simulate the evolution of the $\helthree$ abundance $X(\helthree)$
during the 
pre-main-sequence phase the initial zero-age main-sequence mode assumes 
the $X(\helthree)$ that would have resulted from evolution over a 
specified period $t_{\helthree}$ at constant conditions,
as described by
\citet{Christ1974} but generalized to allow a non-zero initial abundance.
A typical value is $t_{\helthree} = 5 \times 10^7$\,yr.

In the CNO cycle the CN part is assumed to be in nuclear equilibrium
and controlled by the rate of the $\nitfourteen(\hyd, \gamma)\oxyfifteen$
reaction.
In addition, the reactions 
$$\oxysixteen(\hyd, \gamma)\floseventeen(\eplus \nu_{\rm e})
\oxyseventeen(\hyd, \helfour)\nitfourteen
$$
and
$$
\nitfifteen(\hyd, \gamma)\oxysixteen
$$
are included; these play an important role in ensuring the gradual conversion
of $\oxysixteen$ to $\nitfourteen$ and hence increasing the importance
of the CN cycle.
This part of the cycle is assumed to be fully controlled by the
$\oxysixteen(\hyd, \gamma)\floseventeen$ reaction and the branching ratio
between the $\nitfifteen(\hyd, \helfour)\cartwelve$ and
$\nitfifteen(\hyd, \gamma)\oxysixteen$ reactions.


Helium burning has been included in the code using the expressions of
\citet{Angulo1999}, and including also the reaction 
$\cartwelve(\helfour, \gamma)\oxysixteen$.
However, the code is unable to deal with helium ignition in a helium flash.
Thus models with helium burning are restricted to masses in excess of 
$2.3 \Msun$ where ignition takes place in a relatively quiet manner.
Also, as in cases of hydrogen burning (cf.\ Sect.~\ref{sec:conv}),
the treatment of semiconvection
that may be required in this phase has not been implemented.

\subsection{Diffusion and settling}
\label{sec:diffus}

Diffusion and settling are treated in the approximations proposed by
\citet{Michau1993},
with revisions kindly provided by Proffitt.
For completeness I give the complete expressions in Appendix A.
If included, diffusion of heavy elements assumes that all elements
behave as fully ionized $\oxysixteen$;
this is a reasonable approximation in the solar case where the outer 
convection zone is relatively deep, but becomes increasingly questionable
in more massive main-sequence stars.
Here, also, effects of selective radiative levitation should be taken into
account \citep[e.g.,][]{Richer1998, Turcot1998};
there are no current plans to do so in the code.
Various approximations to turbulent diffusion can be included,
partly inspired by \citet{Proffi1991}.
In addition, the code has the option to suppress settling in the outer
parts of the star, to allow modelling of diffusion and settling in the
cores of relatively massive stars where otherwise settling beneath the thin 
outer convection zone would result in a complete depletion of the surface
layers of elements heavier than hydrogen.

At present, diffusion and settling is coupled to nuclear evolution in
the consistent manner of Eq. \ref{eq:dif} only for helium.
For the remaining elements taking part in the nuclear network diffusion 
is neglected.
Correcting this deficiency is an obvious priority.

\section{Treatment of convection}
\label{sec:conv}

The temperature gradient in convection zones is usually computed using the
\citet{Vitens1953} and \citet{Bohm1958} version of mixing-length theory; 
for completeness, the expressions used are provided in Appendix B.
The mixing length is a constant multiple, $\alpha_{\rm ML} H_p$, of the
pressure scale height $H_p$.
In addition, emulations of the \citet{Canuto1991} formulation,
established by \citet{Montei1996}, can be used.

Convective regions can obviously, at least in stars that are not extremely
evolved, be assumed to have uniform composition.
This can in principle be achieved by including a very high diffusion 
coefficient in such regions.
In ASTEC this is used in convective envelopes, ensuring 
that they are chemically uniform.
The treatment of convective cores remains a concern and an area of active
development, however.
Given the
lack of a proper treatment of the diffusion of all elements an explicit
calculation of the chemical evolution is required.
This is characterized by the (assumed homogeneous) abundances $X_{k, \rm c}$
of the elements.
The rate of change of these abundances can be written
\begin{equation}
{\dd X_{k, \rm c} \over \dd t} 
= \bar \CR_k 
+ {1 \over q_{\rm cc}} {\dd q_{\rm cc} \over \dd t} 
[X_k(x_{\rm cc}) - X_{k, \rm c}]
+ {1 \over q_{\rm cc}} \tilde Y_k(x_{\rm cc}) \; ;
\label{eq:ccoreab}
\end{equation}
here $q_{\rm cc}$ is the mass fraction in the convective core,
$x_{\rm cc} = \logten(q_{\rm cc})$, and
\begin{equation}
\bar \CR_k = {1 \over q_{\rm cc}} \int_0^{q_{\rm cc}} \CR_k \dd q 
\end{equation}
is the reaction rate averaged over the convective core.
In the second term in Eq.~\ref{eq:ccoreab} $X_k(x_{\rm cc})$ is
evaluated just outside the core; this term only has an effect if
there is a composition discontinuity at the edge of the core, i.e.,
if the core is growing and diffusion is neglected.
Finally, the term in $\tilde Y_k(x_{\rm cc})$ is obviously only included
in cases where diffusion is taken into account.


In models with a growing convective core, and neglecting diffusion,
a discontinuity is set up in the hydrogen abundance at the edge of the core
(see also Fig.~\ref{fig:rgrad}).
This situation arises in intermediate-mass stars (with masses up to around
$1.7 \Msun$) where the gradual conversion during evolution of 
$\oxysixteen$ into $\nitfourteen$ causes an increase in the importance of
the CNO cycle in the energy generation
(for more massive stars the CNO cycle dominates even with the initial
$\nitfourteen$ abundance).
Since pressure and temperature are obviously continuous, there is also 
a discontinuity in density and opacity $\kappa$, leading also to a jump in the
radiative temperature gradient $\nablarad$ (cf.\ Eq. \ref{eq:radgrad}).
Since $\kappa$ increases with $X$ and $\rho$, while $\rho$ decreases
with increasing $X$, it is not {\it a priori} clear how $\kappa$ and
hence $\nablarad$ react at the discontinuity;
in practice the common behaviour is that $\nablarad$ increases going from
the value of $X$ in the convective core to the higher value in the 
radiative region just outside it.
This raises the question of the definition of convective instability:
if the border of the convective core is defined using the composition of
the core, the region immediately outside the core is therefore convectively
unstable.
As a consequence, ASTEC defines the border of the convective core by the
abundance in the radiative region, leaving a small {\it convectively stable\/}
region just below the border, which nevertheless is assumed to be fully 
mixed in the standard ASTEC implementation.
This may be regarded as an example of semiconvection, of somewhat uncertain
physical consequences \citep[e.g.,][]{Merryf1995}.
A different scheme, now under implementation, 
is discussed in Section~\ref{sec:develop}.

Various options exist for the treatment of convective overshoot.
Simple formulations involve compositionally fully mixed overshoot regions 
from the convective core or convective envelope, over a distance
$\alpha_{\rm ov} H_p$ below a convective envelope, or
$\alpha_{\rm ov} \min( r_{\rm cc} , H_p)$ above a convective core, 
where $r_{\rm cc}$ is the radius of the core.
The overshoot region may be taken to be either adiabatically or radiatively
stratified.
A more elaborate treatment of overshoot from a convective envelope has been
implemented, following \citet{Montei1994},
where various forms of the temperature gradient can be specified,
still assuming the overshoot region to be fully mixed.
This is being extended to emulate the overshoot simulations 
by \citet{Rempel2004}.

\section{Implementation details}
\label{sec:impl}

%
When computing models of the present Sun it is important to adjust 
the input parameters such as to obtain a model that precisely matches
the observed solar radius, luminosity and ratio $Z_{\rm s}/X_{\rm s}$
between the abundances of heavy elements and hydrogen, at the present
age of the Sun.
This is achieved by adjusting the initial hydrogen and heavy-element 
abundances $X_0$ and $Z_0$ as well as the mixing-length parameter
$\alpha_{\rm ML}$ (or another parameter characterizing the treatment of
convection).
In ASTEC the iteration to determine these parameters is carried out
automatically, making the computation of solar models, and the exploration
of the consequences of modifications to the input physics, rather convenient.

The ASTEC code has grown over three decades, with a substantial number of
different uses along the way.
This is clearly reflected in the structure of the code as well as in
the large number of input parameters and flags that control its different
options.
These are provided in an input file, in many cases using simply the
defaults provided in the source of the code.
Also, several different executables can be produced, reflecting partly the
evolution of the code and partly different versions of the physics, 
in particular the equation of state and the opacity, as well as the option
to include diffusion and settling.
To make the code somewhat more user-friendly, scripts have been made which
allow simply to change a few key parameters, such as mass, heavy-element 
abundance and number of timesteps, by editing templates of the input files.
Consequently, the code has been used with success by several users,
including students and postdocs at the University of Aarhus.

In addition to summary output files listing global properties of the models
in the evolution sequence, output of detailed models, on the full
mesh of the calculation, can be made in three
different forms: the so-called {\tt emdl} files, including just the variables
$\{z_i\}$ actually used in the calculation, as well as a complete listing
of the input parameters, to provide full documentation of the calculation;
the {\tt amdl} files which provide the variables needed for the Aarhus
adiabatic pulsation code 
\citep[see][]{Christ2007a};
and the {\tt gong} files, giving an extensive set of variables for use in
further investigations of the models or plotting.
A convenient way to use the code, without overloading storage systems with
the large {\tt gong} files, is to store the full {\tt emdl} file and
subsequently read in models at selected timesteps, to output the corresponding
{\tt amdl} or {\tt gong} files.

For asteroseismology it is evidently crucial to compute oscillation frequencies
of the computed models.
The full calculation of frequencies, for given input parameters to the evolution
calculation, 
often needs to be carried out as part of a larger computation, e.g., when 
fitting observed frequencies to determine the properties of the model.
To facilitate this a version of the code has been made where the evolution
calculation and all parts of the adiabatic oscillation calculation is 
carried out by a single subroutine call, with internal passing of the
intermediate products of the calculation.
This subroutine can then be called by, for example, a fitting code.
An example of such use is the application of the code in
genetic-algorithm fitting \citep[e.g.,][]{Metcal2003},
under development by T. Metcalfe, High Altitude Observatory. 

The code has been implemented on a variety of platforms and appears to
be relatively robust.
To simplify the installation, a complete {\tt tar} package including all the
required files, with a setup script and a full makefile, has been established.
However, the complexity of the code and the lack of adequate documentation
makes it unrealistic to release it for general use.


\section{Further developments}
\label{sec:develop}

From the preceding presentation it is obvious that there are significant 
deficiencies in ASTEC, and work is ongoing to correct them.
A fairly trivial issue is the restricted nuclear network and the failure
to include all elements undergoing nuclear reactions in the full diffusive
treatment.
Rather more serious problems concern the treatment of the borders of
convective regions;
even though this obviously also involves open issues of a basic physical
nature, the code should at least aim at treating these regions in a
numerically consistent, even if perhaps not physically adequate, manner.
A serious problem is the failure of the code for models with convective 
cores, when diffusion and settling of both helium and heavy elements are
included \citep[cf.][]{Christ2007b};
on the other hand, the case of just helium diffusion can be handled.
This problem may be related to issues of semiconvection where convective
stability is closely related to the details of the composition profile
\citep{Montal2007}.

\begin{figure}
\begin{center}
\includegraphics[width=8.6cm]{\fig/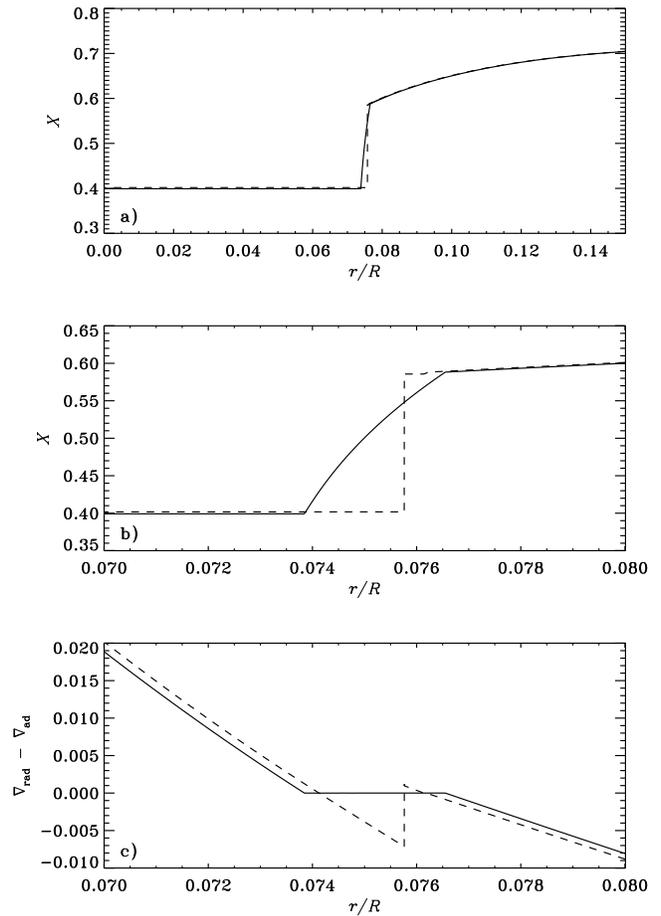}
\end{center}
\caption{
Hydrogen-abundance profile and $\nablarad - \nablaad$ against fractional radius
in two $1.5 \Msun$ models of age 1.36 Gyr and with $Z = 0.02$.
The dashed curves illustrate the usual treatment in ASTEC, where the boundary
of the convective core is determined by the composition in the radiative 
region just outside it, and the hydrogen abundance $X$ is discontinuous.
In the model illustrated by the solid curves a gradient in $X$ is set up such
that $\nablarad - \nablaad = 0$ in the outer parts of the convective core.
}
\label{fig:rgrad}       
\end{figure}

In ASTEC I have considered two cases of semiconvection, although the 
implementation is still under development.
One case concerns growing convective cores in non-diffusing models
(see also Sect.~\ref{sec:conv}),
as illustrated in Fig.~\ref{fig:rgrad}.
The dashed curves illustrate the properties of a model computed with the
normal ASTEC implementation, where the hydrogen abundance $X$ is discontinuous
at the boundary of the core (see panels a and b).
As discussed above, the extent of the convective core is determined by
the behaviour of the radiative temperature gradient $\nablarad$ in the
radiative region (see panel c),%
\footnote{Owing to slight convergence problems in this calculation,
$\nablarad - \nablaad$ is in fact positive at the point identified as
the convective boundary.}
leading to a convectively stable region just beneath
the boundary which nonetheless is assumed to be fully mixed.
A possibly more reasonable treatment, illustrated by the continuous curves,
assumes that a hydrogen-abundance profile is established such that
$\nablarad = \nablaad$ in the outermost parts of the convective core.
Since this affects a very small part of the core of the model the effects
on its global properties are modest;
however, it may have some influence on its pulsational properties,
particularly for g modes or interface modes predominantly trapped near
the boundary of the core, possibly offering the potential for 
asteroseismic diagnostics of this behaviour.
Significant sensitivity to the treatment of this region was
indeed found by \citet{Moya2007}; a similar model,
but computed without such treatment of semiconvection (and with somewhat 
inadequate mesh resolution of the critical region) displayed far larger
frequency differences
than did the model considered in Fig.~\ref{fig:rgrad}
between oscillation codes using different numerical techniques,
particularly for modes
with g-mode like behaviour and partly trapped near the edge of the core.

The second aspect of semiconvection concerns the base of the convective 
envelope in models with diffusion and settling of helium and heavy elements.
As noted by \citet{Bahcal2001} and discussed in more detail by
\citet{Christ2007c} the gradient in the
heavy-element abundance $Z$ established by settling beneath the convection zone
leads to convective instability, e.g., in $1 \Msun$ models of age somewhat
higher than the present Sun.
Complete mixing of the unstable region removes the gradient and hence the
cause of the instability, leading to an uncertain situation, characteristic
of semiconvection.
As above, a perhaps reasonable assumption might be that partial mixing takes
place to establish composition gradients that ensure neutral convective
stability in the critical region, with $\nablarad = \nablaad$.
Since the opacity depends both on $X$ and $Z$ the mixing must consistently
affect both abundances.
I have attempted to implement this by including a turbulent diffusivity,
obviously common to all elements, determined iteratively as a function
of depth beneath the convection zone such that 
the resulting profiles of $X$ and $Z$ lead to neutral stability.

\section{Concluding remarks}

No stellar evolution code is probably ever finished or fully tested.
ASTEC has certainly proved useful in a number of applications, 
and the results for the Sun, as applied to helioseismology, are perhaps
reasonably reliable, at least within the framework 
of `standard solar modelling'.
It is obvious, however,
that application to the increasingly accurate and detailed asteroseismic data
will require further development.
The tests provided through the ESTA collaboration and extended through the
HELAS Coordination Action are certainly most valuable in this regard.

\begin{acknowledgements}
I am very grateful to D. O. Gough for his assistance in developing the 
initial version of the code, including the basic package to solve the 
equations of stellar evolution.
Many people have contributed to the development of ASTEC over the years,
and I am very grateful for their contributions.
I thank W. D\"appen for contributing the MDI and OPAL equation-of-state
packages, G. Houdek for providing the opacity interpolation packages and
the OPAL data in the appropriate form,
M. J. P. F. G. Monteiro for providing alternate treatments of 
the parametrization of convection,
C. R. Proffitt for help with
installing diffusion and settling in the code and M. Bazot for assistance 
with updating the treatment of nuclear reactions.
M. J. P. F. G Monteiro is thanked for organizing the ESTA effort and 
Y. Lebreton and J. Montalb\'an for taking care of the stellar evolution part of
it.
This project is being supported by
the Danish Natural Science Research Council and by
the European Helio- and Asteroseismology Network (HELAS),
a major international collaboration funded by the European Commission's
Sixth Framework Programme.
\end{acknowledgements}

%

\appendix

\section{Treatment of diffusion and settling}

For completeness I reproduce the expressions used to compute the diffusion
and settling coefficients.
These are largely obtained from \citet{Michau1993}, although with minor
modifications.

In terms of the variable $\tilde Y_k$ introduced in Section~\ref{sec:eq}
Eqs \ref{eq:difx} and \ref{eq:dify} become
\begin{eqnarray}
{\partial X_k \over \partial x} & = &
{\CM M q \over \CD_k} (M \tilde Y_k - \CV_k X_k) \; , 
\label{eq:difxa} \\
{\partial \tilde Y_k \over \partial x} & = &
\CM q  \left({\partial X_k \over \partial t} - \CR_k \right) \; .
\label{eq:difya}
\end{eqnarray}

For hydrogen ($k= 1$)
\begin{equation}
\CV_1 = -{4 \pi B T^{5/2} \over \log \Lambda_{\rm HHe} (0.7+0.3X)}
\left({5 \over 4} + {9 \over 8} \nabla \right) {G \rho m \over p}
(1 - X - Z) \; ,
\end{equation}
where 
\begin{equation} 
B = {15 \over 16} \left( {2 m_{\rm u} \over 5 \pi} \right)^{1/2}
{k_{\rm B}^{5/2} \over e^4} \; ,
\end{equation}
$m_{\rm u}$ being the atomic mass unit, $k_{\rm B}$ Boltzmann's constant
and $e$ the electron charge.
Also, $\log \Lambda_{\rm HHe}$ is the Coulomb logarithm, approximated by
\begin{equation}
\log \Lambda_{\rm HHe} = -19.95 - {1 \over 2} \log \rho 
- {1 \over 2} \log \left({ X + 3 \over 2}\right) +{3 \over 2} \log T \; .
\end{equation}
In the above expressions, cgs units are used and `$\log$' is natural logarithm.
The diffusion coefficient is given by
\begin{equation}
\CD_1 = (4 \pi \rho r^2)^2 {B T^{5/2} \over \rho \log \Lambda_{\rm HHe} 
(0.7 + 0.3X)} { (3+X) \over (1+X)(3+5X)} \; .
\end{equation}

For a trace element $k$, with charge $Z_k$ and atomic mass $A_k$,
\begin{equation}
\CD_k = (4 \pi \rho r^2)^2 { 2 B T^{5/2} \over 5^{1/2} \rho Z_k^2}
{1 \over (\CC_k + C_{k\,\rm He} A_{k\,\rm He}^{1/2})} \; ,
\end{equation}
where 
\begin{equation}
\CC_k = X( A_{k\,\rm H}^{1/2} C_{k\,\rm H}
- A_{k\,\rm He}^{1/2} C_{k\,\rm He}) \; .
\end{equation}
Here $C_{k\,\rm H}$ is the collision integral between element $k$ and
hydrogen, which \citet{Michau1993} approximate, fitting numerically
determined values, as
\begin{equation}
C_{k\,\rm H} = {1 \over 1.2} \log [ \exp(1.2 \log \Lambda_{k\,\rm H})
+1 ] \; ,
\end{equation}
where
\begin{equation}
\Lambda_{k\,\rm H} = \Lambda_{\rm H He} - \log Z_k + \log 2 \; ;
\end{equation}
similarly the collision integral $C_{k\,\rm H}$ between element $k$ and
helium is evaluated as
\begin{equation}
C_{k\,\rm He} = {1 \over 1.2} \log [ \exp(1.2 \log \Lambda_{k\,\rm He})
+1 ] \; ,
\end{equation}
where
\begin{equation}
\Lambda_{k\,\rm He} = \Lambda_{\rm H He} - \log Z_k \; .
\end{equation}
Also, 
\begin{equation}
A_{k\,\rm H} = {A_k A_{\rm H} \over A_k + A_{\rm H}}
\end{equation}
is the reduced mass in the element $k$, hydrogen system,
$A_{\rm H}$ being the atomic mass of hydrogen;
the reduced mass $A_{k, \rm He}$ involving helium is defined similarly.

As presented by \citet{Michau1993} the diffusion velocity of 
trace elements depends on the gradient in the hydrogen abundance.
To incorporate this in the formalism described by
Eqs \ref{eq:difxa} and \ref{eq:difya}, I express $\partial X / \partial r$
in terms of $\tilde Y_1$ and write the coefficient $\CV_k$ as
\begin{equation}
\CV_k = \CV_k^{(1)} + \CV_k^{(2)} M \tilde Y_1 \; .
\end{equation}
Here
\begin{eqnarray}
&& \CV_k^{(1)} = 
 -4 \pi B T^{5/2} \rho \left\{
{2 [ 1 + Z_k - A_k(5X + 3)/4] \over
5^{1/2} Z_k^2 (\CC_k + C_{k\,\rm He} A_{k\,\rm He}^{1/2})} \right. \\
&& \qquad\qquad \left. - {0.54 (4.75 X + 2.25) \over 
\log \Lambda_{\rm HHe} + 5} \nabla
\right\} {G m \over p } \nonumber \\
&& \qquad + {\CV_1 \over \CD_1} (4 \pi r^2 \rho)^2
{2 B T^{5/2} \over 5^{1/2} \rho Z_k^2 
(\CC_k + C_{k\,\rm He} A_{k\,\rm He}^{1/2})}
{2 Z_k + 5(1+X) \over (1+X)(5X+3)}X \; , \nonumber
\end{eqnarray}
and
\begin{eqnarray}
&& \CV_k^{(2)} = \\
&& -{(4 \pi r^2 \rho)^2 \over \CD_1}
{2 B T^{5/2} \over 5^{1/2} \rho Z_k^2 }
{2 Z_k + 5(1+X) \over (1+X)(5X+3)}
+ {\CC_k \over X (\CC_k + C_{k\,\rm He} A_{k\,\rm He}^{1/2})} - 0.23 \; .
\nonumber
\end{eqnarray}

\section{Mixing-length formulation}

%
The calculation of 
$\nabla$ in convective regions is carried out using the \citet{Vitens1953}
mixing length theory, in the form given by \citet{Gough1977}.
This is expressed in terms of
\begin{equation}
\CR = {g \delta \ell^4 \over (K/\rho c_p)^2 H_p} \; ,
\label{eq:mixlen2}
\end{equation}
\begin{equation}
A = 2 \eta^{-1} \CR^{-1/2} ( \nablarad - \nablaad)^{-1/2} \; ,
\label{eq:mixlen3}
\end{equation}
and
\begin{equation}
\lambda = {4 \over 3 \sqrt{2}} \eta \left({\Phi \over 2} \right)^{1/2} \; .
\label{eq:mixlen4}
\end{equation}
Here $H_p = p/(\rho g)$ is the pressure scale height, $g = Gm/r^2$ is the
gravitational acceleration, 
$\delta = - (\partial \log \rho / \partial \log T)_p$,
$K = 4 a c T^3/(3 \kappa \rho)$ is the radiative conductivity
and $c_p$ is the specific heat at constant pressure.
Also $\ell$ is the mixing length, which is taken to be a constant
multiple of the pressure scale height, $\ell = \alpha_{\rm ML} H_p$,
and $\eta$ and $\Phi$ are geometrical quantities, assumed be constant, 
which are related to the aspect ratio of the convective cells; for
$\eta = \sqrt{2}/9$ and $\Phi = 2$ we get the \citet{Bohm1958} expressions
(these values are used in the computation).
Then
\begin{equation}
\nabla - \nablaad = Y(Y + A) (\nablarad - \nablaad) \; ,
\label{eq:mixlen5}
\end{equation}
where $Y$ is the positive root of
\begin{equation}
{1 \over 3 \lambda A} Y^3 + Y^2 + A Y - 1 = 0 \; .
\label{eq:mixlen6}
\end{equation}
The general solution to Eq.~\ref{eq:mixlen6} is
\begin{equation}
Y = A \left[ {x \over A} - \lambda(1 - \lambda) {A \over x} 
- \lambda \right] \; ,
\label{eq:mixlen7}
\end{equation}
where
\begin{equation}
x = A \left\{\gamma + [\gamma^2 + \lambda^3(1 - \lambda)^3]^{1/2}
\right\}^{1/3} \; ,
\label{eq:mixlen8}
\end{equation}
and
\begin{equation}
\gamma = \lambda \left[ {3 \over 2 A^2} + 
\lambda \left( {3 \over 2} - \lambda \right) \right] \; .
\label{eq:mixlen9}
\end{equation}
For large and small $A$ asymptotic expressions for $Y$ are easily found.
For $A \gg 1$
\begin{equation}
Y \simeq A^{-1} \left[1 - A^{-2} + \left(2 - {1 \over 3 \lambda} \right)
A^{-4} \right] \; ,
\label{eq:mixlen10}
\end{equation}
and for $A \ll 1$
\begin{equation}
Y \simeq (3 \lambda A)^{1/3} \left[ 1 - {1 \over 3} (3 \lambda A)^{2/3} \right] \; .
\label{eq:mixlen11}
\end{equation}
Equation \ref{eq:mixlen10} is used when $A \ge 15$ and Eq.~\ref{eq:mixlen11}
when $A \le 10^{-5}$.
At these points
the relative differences between the asymptotic values of $Y$ and
those found from Eq.~\ref{eq:mixlen7} are less than $5 \times 10^{-7}$.

\end{document}